\documentclass{llncs}

\title{A Comparative Study of Arithmetic Constraints on Integer Intervals}

\author{Krzysztof R. Apt\inst{1,2} \and Peter Zoeteweij\inst{1}}
\institute{CWI, P.O. Box 94079, 1090 GB Amsterdam, the Netherlands\\
\and
University of Amsterdam, the Netherlands}

 \newcommand{\eclipse}{ECL$^i$PS$^e$}
\newcommand{\Proof}{\NI
                    {\bf Proof.}\ }
 \usepackage{amsmath}
 \usepackage{latexsym}
 \usepackage{alltt}

\newcommand{\LL}{\mbox{$\ldots$}}
\newcommand{\oldbfe}[1]{\begin{bfseries}\emph{#1}\end{bfseries}}
\newcommand{\p}[2]{\langle #1 \ ; \ #2 \rangle}
\newcommand{\C}[1]{\mbox{$\{{#1}\}$}}           % curly braces
\newcommand{\ES}{\mbox{$\emptyset$}}
\newcommand{\te}{\mbox{$\exists$}}
\newcommand{\HB}{\hfill{$\Box$}}
\newcommand{\II}{\vspace{2 mm}}
\newcommand{\NI}{\noindent}
\newcommand{\adjceiling}[1]{\left \lceil #1 \right \rceil}
\newcommand{\adjfloor}[1]{\left \lfloor #1 \right \rfloor}
\newcommand{\sse}{\mbox{$\:\subseteq\:$}}
\newcommand{\VV}{\vspace{5 mm}}

\begin{document}

\date{}
\maketitle

\begin{abstract}
We propose here a number of approaches to implement constraint
  propagation for arithmetic constraints on integer intervals.  To
  this end we introduce integer interval arithmetic.  Each approach is
  explained using appropriate proof rules that reduce the variable
  domains. We compare these approaches using a set of benchmarks.
\end{abstract}

\section{Preliminaries}
\label{sec:preliminaries}
\subsection{Introduction}

The subject of arithmetic constraints on reals has attracted a
great deal of attention in the literature. For some reason arithmetic
constraints on integer intervals have not been studied even though
they are supported in a number of constraint programming systems. In
fact, constraint propagation for them is present in \eclipse{},
SICStus Prolog, GNU Prolog, ILOG Solver and undoubtedly most of the
systems that support constraint propagation for linear constraints on
integer intervals.  Yet, in contrast to the case of linear
constraints --- see notably \cite{HS03} ---
 we did not encounter in the literature any analysis of
this form of constraint propagation.  

In this paper we study these constraints in a systematic
way.  It turns out that in contrast to linear constraints on integer
intervals there are a number of natural approaches to
constraint propagation for these constraints.  

To define them we introduce integer interval arithmetic that is
modeled after the real interval arithmetic, see e.g., \cite{HJvE01}.
There are, however, essential differences since we deal with integers
instead of reals. For example, multiplication of two integer intervals
does not need to be an integer interval.  In passing by we show that
using integer interval arithmetic we can also define succinctly the
well-known constraint propagation for linear constraints on integer intervals.
In the second part of the paper we compare the proposed approaches by
means of a set of benchmarks.

\subsection{Constraint Satisfaction Problems}

We review here the standard concepts of a constraint and of a
constraint satisfaction problem.  Consider a sequence of variables $X
:= x_1, \LL, x_n$ where $n \geq 0$, with respective domains $D_1, \LL,
D_n$ associated with them.  So each variable $x_i$ ranges over the
domain $D_i$.  By a \oldbfe{constraint} $C$ on $X$ we mean a subset of
$D_1 \times \LL \times D_n$.  Given an element $d := d_1, \LL, d_n$ of
$D_1 \times \LL \times D_n$ and a subsequence $Y := x_{i_1}, \LL,
x_{i_\ell}$ of $X$ we denote by $d[Y]$ the sequence $d_{i_1}, \LL,
d_{i_{\ell}}$. In particular, for a variable $x_i$ from $X$, $d[x_i]$
denotes $d_i$.

A \oldbfe{constraint satisfaction problem}, in short CSP, consists of
a finite sequence of variables $X$ with respective domains ${\cal
  D}$, together with a finite set $\cal C$ of constraints, each on a
subsequence of $X$. We write it as $\p{{\cal C}}{x_1 \in D_1,
  \LL, x_n \in D_n}$, where $X := x_1, \LL, x_n$ and ${\cal D} :=
D_1, \LL, D_n$.

By a \oldbfe{solution} to $\p{{\cal C}}{x_1 \in D_1, \LL, x_n \in D_n}$
we mean an element $d \in D_1 \times \LL \times D_n$ such that for
each constraint $C \in {\cal C}$ on a sequence of variables $X$ we
have $d[X] \in C$.  We call a CSP \oldbfe{consistent} if it has a
solution and \oldbfe{inconsistent} if it does not.
Two CSPs with the same sequence of variables are called
\oldbfe{equivalent\/} if they have the same set of solutions.  In what
follows we consider CSPs the constraints of which are defined in a
simple language and identify the syntactic description of a constraint
with its meaning being the set of tuples that satisfy it.

We view \oldbfe{constraint propagation} as a process of transforming CSPs that maintains
their equivalence. In what follows we define this process by means
of proof rules that act of CSPs and preserve equivalence.
An interested reader can consult \cite{Apt98a} for a precise explanation of 
this approach to describing constraint propagation.

\subsection{Arithmetic Constraints}

To define the arithmetic constraints use the alphabet that comprises
\begin{itemize}
\item variables,

\item two constants, 0 and 1,

\item the unary minus function symbol `$-$',
  
\item three binary function symbols,  `+',`$-$'and `$\cdot$', all written in the
  infix notation.
\end{itemize}

By an \oldbfe{arithmetic expression} we mean a term formed in this
alphabet and by an \oldbfe{arithmetic constraint} 
a formula of the form
\[
s \: op \: t,
\]
where $s$ and $t$ are arithmetic expressions and $op \in \C{<, \leq, =, \neq, \geq, >}$.
For example
\begin{equation}
x^5 \cdot y^2 \cdot z^4  + 3 x \cdot y^3 \cdot z^5 \leq 10 + 4 x^4 \cdot y^6 \cdot z^2 - y^2 \cdot x^5 \cdot z^4
\label{eq:arithcon}
\end{equation}
is an arithmetic constraint. Here $x^5$ is an abbreviation for $x \cdot x \cdot x \cdot x \cdot x$ and similarly
with the other expressions.
If `$\cdot$' is not used in an arithmetic constraint, we call it a \oldbfe{linear constraint}.

By an \oldbfe{extended arithmetic expression} we mean a term formed in the above
alphabet extended by the unary function symbols `$.^n$' and `$\sqrt[n]{.}$' for each $n \geq 1$ and
the binary function symbol `$/$' written in the infix notation.
For example
\begin{equation}
\sqrt[3]{(y^2 \cdot z^4)/(x^2 \cdot u^5)}
\label{eq:extended}
\end{equation}
is an extended arithmetic expression. Here, in contrast to the above
$x^5$ is a term obtained by applying the function symbol `$.^5$' to the
variable $x$.
The extended arithmetic expressions will be used only to define constraint
propagation for the arithmetic constraints.

Fix now some arbitrary linear ordering $\prec$ on the variables of the language.
By a \oldbfe{monomial} we mean an integer or a term
of the form
\[
a \cdot x_{1}^{n_1} \cdot \LL   \cdot x_{k}^{n_k}
\]
where $k > 0$, $x_1, \LL, x_k$ are different variables ordered w.r.t.~
$\prec$, and $a$ is a non-zero integer and $n_1, \LL, n_k$ are
positive integers.  We call then $x_{1}^{n_1} \cdot \LL \cdot
x_{k}^{n_k}$ the \oldbfe{power product} of this monomial.

Next, by a \oldbfe{polynomial} we mean a term of the form
\[
\Sigma^{n}_{i=1} m_i,
\]
where $n > 0$, at most one monomial $m_i$ is an integer, and the power
products of the monomials $m_1, \LL, m_n$ are pairwise different.
Finally, by a \oldbfe{polynomial constraint} we mean an arithmetic
constraint of the form $s \: op \: b$, where $s$ is a polynomial with
no monomial being an integer, $op \in \C{<, \leq, =, \neq, \geq, >}$,
and $b$ is an integer.  It is clear that
by means of appropriate transformation rules we can transform each
arithmetic constraint to a polynomial constraint. For example,
assuming the ordering $x \prec y \prec z$ on the variables, the
arithmetic constraint (\ref{eq:arithcon}) can be transformed to the
polynomial constraint
\[
2x^5 \cdot y^2 \cdot z^4 - 4 x^4 \cdot y^6 \cdot z^2  + 3 x \cdot y^3 \cdot z^5 \leq 10
\]
So, without loss of generality, from now on we shall limit our
attention to the polynomial constraints.

Next, let us discuss the domains over which we interpret the arithmetic constraints.
By an \oldbfe{integer interval}, or an \oldbfe{interval} 
in short, we mean
an expression of the form
\[
[a..b]
\]
where $a$ and $b$ are integers;
$[a..b]$ denotes the set of all integers between $a$ and $b$,
including $a$ and $b$.
If $a > b$, we call $[a..b]$ the \oldbfe{empty interval} and denote it by $\ES$.
Finally, by a \oldbfe{range} we mean an expression of the form
\[
x \in I
\]
where $x$ is a variable and $I$ is an interval.

\section{Integer Set Arithmetic}
\label{sec:interval}

To reason about the arithmetic constraints we employ a generalization of the
arithmetic operations to the sets of integers.

\subsection{Definitions}

For $X,Y$ sets of integers we define the following
operations:

\begin{itemize}

\item addition:
\[
X + Y := \C{x + y \mid x \in X, y \in Y},
\]

\item subtraction:
\[
X - Y := \C{x - y \mid x \in X, y \in Y},
\]

\item multiplication:
\[
X \cdot Y := \C{x \cdot y \mid x \in X, y \in Y},
\]

\item division:
\[
X/Y  := \C{u \in {\cal Z} \mid \te x \in X \te y \in Y \: u \cdot y = x},
\]

\item exponentiation:
\[
X ^ n := \C{x ^n \mid x \in X},
\]
for each natural number $n > 0$,

\item root extraction:
\[
\sqrt[n]{X} := \C{x \in {\cal Z} \mid x ^n \in X},
\]
for each natural number $n > 0$.

\end{itemize}

All the operations except division are defined in the expected way.
We shall return to it at the end of Section \ref{sec:third}.  At the
moment it suffices to note the division operation is defined for all
sets of integers, including $Y = \ES$ and $Y = \C{0}$.
This division operation corresponds to the following division operation on
the sets of reals introduced in \cite{Rat96}: 

\[
\frac{X}{Y}  := \C{u \in {\cal R} \mid \te x \in X \te y \in Y \: u \cdot y = x}.
\]
For a (n integer or real) number $a$ and $op \in \C{+, -, \cdot, / }$ we identify
$a \: op \: X$ with $\C{a} \: op \: X$ and $X \: op \: a$ with $X \: op \: \C{a}$.

To present the rules we are interested in we shall also use the
addition and division operations on the sets of real numbers.
Addition is defined in the same way as for the sets of integers, and
division is defined above. In \cite{HJvE01} it is explained
how to implement these operations.

Further, given a set $A$ of integers or reals, we define
\[
^{\leq}A := \C{x \in {\cal Z} \mid \te a \in A \: x \leq a},
\]
\[
^{\geq}A := \C{x \in {\cal Z} \mid \te a \in A \: x \geq a}.
\]

When limiting our attention to intervals of integers the
following simple observation is of importance.  

\begin{note} \label{note:1}
For $X,Y$ integer intervals and $a$ an integer the following holds:

\begin{itemize}
\item $X \cap Y$, $X + Y, X - Y$ are integer intervals.
\item $X/\C{a}$ is an integer interval.
\item $X \cdot Y$ does not have to be an integer interval, even if $X = \C{a}$ or $Y = \C{a}$.
\item $X/Y$ does not have to be an integer interval. 
\item For each $n >1$ \ $X ^ n$ does not have to be an integer interval.
\item For odd $n >1$ \ $\sqrt[n]{X}$ is an integer interval.
\item For even $n >1$ \ $\sqrt[n]{X}$ is an integer interval or a disjoint union of two integer intervals.
\HB
\end{itemize}
\end{note}

For example we have
\[
[2..4] + [3..8] = [5..12],
\]
\[
[3..7] - [1..8] = [-5..6],
\]
\[
[3..3] \cdot [1..2] = \C{3,6},
\]
\[
[3..5]/[-1..2] = \C{-5,-4,-3,2,3,4,5},
\]
\[
[-3..5]/[-1..2] = {\cal Z},
\]
\[
[1..2]^2 = \C{1,4},
\]
\[
\sqrt[3]{[-30..100]} = [-3..4],
\]
\[
\sqrt[2]{[-100..9]} = [-3..3],
\]
\[
\sqrt[2]{[1..9]} = [-3..-1] \cup [1..3].
\]
To deal with the problem that non-interval domains can be produced
by some of the operations we introduce the following operation 
on the subsets of the set of the integers ${\cal Z}$:
\[
int(X) :=    
\left\{ 
\begin{tabular}{ll}
\mbox{smallest integer interval containing $X$} &  \mbox{if $X$ is finite,} \\
${\cal Z}$                                   &  \mbox{otherwise.}
\end{tabular}
\right . 
\]
For example $int([3..5]/[-1..2]) = [-5..5]$
and $int([-3..5]/[-1..2]) = {\cal Z}$.

\subsection{Implementation}

\label{subsec:implementation}

To define constraint propagation for the arithmetic constraints on integer intervals
we shall use the integer set arithmetic, mainly limited to the integer intervals.
This brings us to the discussion of how to
implement the introduced operations on the integer intervals.
Since we are only interested in maintaining the property that the sets remain integer
intervals or the set of integers ${\cal Z}$
we shall clarify how to implement the
intersection, addition, subtraction and root extraction operations of the integer intervals and the
$int(.)$ closure of the multiplication, division and exponentiation operations on the
integer intervals.  The case when one of the intervals is empty is easy to deal with. So we
assume that we deal with non-empty intervals
$[ a..b ]$ and $[ c..d ]$, that is $a \leq b$ and $c \leq d$.

\paragraph{Intersection, addition and subtraction}
It is easy to see that
\[
[ a..b ] \cap [ c..d ] = [ max(a,c)..min(b,d)],
\]
\[
[ a..b ] +[ c..d ] = [a+c\ ..\ b+d],
\]
\[
[ a..b ] - [ c..d ] = [a-d\ ..\ b-c].
\]
So the interval intersection, addition, and subtraction are straightforward to implement.

\paragraph{Root extraction}

The outcome of the root extraction operator applied to an integer
interval will be an integer interval or a disjoint union of two
integer intervals.  We shall explain in Section \ref{sec:first} why it is
advantageous not to apply $int(.)$ to the outcome.  This operator can
be implemented by means of the following case analysis.  
\II

\NI
\emph{Case 1.} Suppose $n$ is odd. Then
\[
\sqrt[n]{[a..b]} = [\adjceiling{\sqrt[n]{a}} .. \adjfloor{\sqrt[n]{b}}].
\]

\NI
\emph{Case 2.} Suppose $n$ is even and $b < 0$. Then

\[
\sqrt[n]{[a..b]} = \ES.
\]

\NI
\emph{Case 3.} Suppose $n$ is even and $b \geq 0$. Then
\[
\sqrt[n]{[a..b]} = [-\adjfloor{|\sqrt[n]{b}|} .. -\adjceiling{|\sqrt[n]{a^{+}}|}]
              \cup [\adjceiling{|\sqrt[n]{a^{+}}|} .. \adjfloor{|\sqrt[n]{b}|} ]
\]
where $a^{+} := \textit{max\/}(0,a)$.

\paragraph{Multiplication}
For the remaining operations we only need to explain how to implement
the $int(.)$ closure of the outcome.
First note that \[
int([ a..b] \cdot [ c..d])   =  [ min(A)..  max(A) ],  
\]
where $A = \C{a \cdot c, a \cdot d, b \cdot c, b \cdot d}$.  

Using an appropriate case analysis we can actually compute the bounds
of $int([ a..b] \cdot [ c..d])$ directly in terms of the bounds of the
constituent intervals.

\paragraph{Division}
In contrast, the $int(.)$ closure of the interval division is not so
straightforward to compute. The reason is that, as we shall see in a
moment, we cannot express the result in terms of some simple
operations on the interval bounds.

Consider non-empty integer intervals $[ a..b] $ and $[ c..d ] $.
In analyzing the outcome of $int([a..b]/[c..d])$ we distinguish the following
cases.
\II

\NI
\emph{Case 1.} Suppose $0 \in [a .. b]$ and $0 \in [c .. d]$.

Then by definition $int([a..b]/[c..d]) = {\cal Z}$.
For example, 
\[
int([-1..100]/[-2..8]) = {\cal Z}.
\]

\NI
\emph{Case 2.} Suppose $0 \not \in [a .. b]$ and $c = d = 0$.

Then by definition $int([a .. b]/[c .. d]) = \ES$.
For example, 
\[
int([10..100]/[0..0]) = \ES.
\]

\NI
\emph{Case 3.} Suppose $0 \not \in [a .. b]$ and $c < 0$ and $0 < d$.

It is easy to see that then
\[
int([a .. b]/[c .. d]) = [-e .. e],
\] 
where $e = max (|a|, |b|)$.
For example, 
\[
int([-100..-10]/[-2..5]) = [-100..100].
\]

\NI
\emph{Case 4.} Suppose $0 \not \in [a .. b]$ and either $c = 0$ and $d \neq 0$ or $c \neq 0$ and $d = 0$.

Then $int([a .. b]/[c .. d]) = int([a .. b]/([c .. d] -  \C{0}))$.
For example 
\[
int([1..100]/[-7..0]) = int([1..100]/[-7..-1]).
\]
This allows us to reduce this case to Case 5 below.

\NI
\emph{Case 5.} Suppose $0 \not \in [c .. d]$.

This is the only case when we need to compute $int([a .. b]/[c .. d])$
indirectly. First, observe that we have 
\[
int([a..b]/[c..d]) \sse [\adjceiling{min(A)} .. \adjfloor{max(A)}],    
\]
where $A = \C{a/c, a/d, b/c, b/d}$.

However, the equality does not need to hold here.  Indeed, note for
example that $int([155..161]/[9..11]) = [16..16]$, whereas for $A =
\{155/9, 155/11, 161/9$, $161/11\}$ we have $\adjceiling{min(A)} = 15$
and $\adjfloor{max(A)} = 17$.  The problem is that the value 16 is
obtained by dividing 160 by 10 and none of these two values is an
interval bound.  

This complication can be solved by preprocessing the interval $[c..d]$
so that its bounds are actual divisors of an element of $[a..b]$.
First, we look for the least $c' \in [c..d]$ such
that $\exists x \in [a..b]\ \exists u \in {\cal Z}\ u\cdot c' = x$.
Using a case analysis, the latter property can be established without search.
Suppose for example that $a>0$ and $c>0$. In this case, if $c' \cdot
\lfloor \frac{b}{c'} \rfloor \geq a$, then $c'$ has the required
property. Similarly, we look for the largest 
$d' \in [c..d]$ for which an analogous condition holds. Now
$\textit{int\/}([a..b] / [c..d]) = [\lceil\textit{min\/}(A)\rceil ..
\lfloor\textit{max\/}(A)\rfloor]$, where $A = \{a/c', a/d', b/c',
b/d'\}$.

\paragraph{Exponentiation}
The $int(.)$ closure of the interval exponentiation is straightforward 
to implement by distinguishing the following cases.
\II

\NI
\emph{Case 1.} Suppose $n$ is odd. Then
\[
int([a..b]^n) = [a^n ..\: b ^n]. 
\]

\NI
\emph{Case 2.} Suppose $n$ is even and $0 \leq a$. Then
\[
int([a..b]^n) = [a^n ..\: b ^n]. 
\]

\NI
\emph{Case 3.} Suppose $n$ is even and $b \leq 0$. Then
\[
int([a..b]^n) = [b^n ..\: a^n]. 
\]

\NI
\emph{Case 4.} Suppose $n$ is even and $a < 0$ and $0 < b$. Then
\[
int([a..b]^n) = [ 0..  max(a^n, b^n) ].
\]

\subsection{Correctness Lemma}

Given now an extended arithmetic expression $s$ each variable of
which ranges over an integer interval, we define $int(s)$ as the
integer interval or the set ${\cal Z}$ obtained by systematically
replacing each function symbol by the application of the $int(.)$
operation to the corresponding integer set operation.  For example,
for the extended arithmetic expression $s := \sqrt[3]{(y^2 \cdot
  z^4)/(x^2 \cdot u^5)}$ of (\ref{eq:extended}) we have
\[
int(s)= int(\sqrt[3]{int(int(Y^2) \cdot int(Z^4))/int(int(X^2) \cdot int(U^5))}),
\]
where $x$ ranges over $X$, etc.

The discussion in the previous subsection shows how to compute
$int(s)$ given an extended arithmetic expression $s$ and the integer
interval domains of its variables.

The following lemma is crucial for our considerations. It is a counterpart of the
so-called `Fundamental Theorem of Interval Arithmetic' established in \cite{Moo66}.
Because we deal here with the integer domains an additional assumption is
needed to establish the desired conclusion.

\begin{lemma}[Correctness] \label{lem:correctness}
Let $s$ be an extended arithmetic expression with the variables
$x_1, \LL, x_n$. Assume that each variable $x_i$ of $s$
ranges over an integer interval $X_i$.
Choose $a_i \in X_i$ for $i \in [1..n]$ and 
denote by $s(a_1, \LL, a_n)$ the result of replacing in $s$
each occurrence of a variable $x_i$ by $a_i$.

Suppose that each subexpression of 
$s(a_1, \LL, a_n)$ evaluates to an integer. Then
the result of evaluating $s(a_1, \LL, a_n)$ is an element of $int(s)$.
\end{lemma}
\Proof The proof follows by a straightforward induction on the structure of $s$.

\HB

\section{An Intermezzo: Constraint Propagation for Linear Constraints}
\label{sec:intermezzo}

Even though we focus here on arithmetic constraints on integer intervals, it is helpful to
realize that the integer interval arithmetic is also useful to define in a succinct way
the well-known rules for constraint propagation for linear constraints.
To this end consider first a constraint $\Sigma_{i=1}^{n} a_i \cdot x_i = b$, where
$n \geq 0$, $a_1, \LL ,a_n$ are non-zero integers, $x_1, \LL ,x_n$ are
different variables, and $b$ is an integer.
To reason about it we can use the following rule parametrized by $j \in [1..n]$:

\begin{center}
\emph{LINEAR EQUALITY}
\[
\frac
{\p{\Sigma_{i=1}^{n} a_i \cdot x_i = b}{x_1 \in D_1, \LL, x_n \in D_n}}
{\p{\Sigma_{i=1}^{n} a_i \cdot x_i = b}{x_1 \in D'_1, \LL, x_n \in D'_n}}
\]
\end{center}
where 

\begin{itemize}
\item for $i \neq j$
\[
D'_i := D_i,
\]

\item 
\[
D'_j := D_j \cap int\Big{(}(b - \Sigma_{i \in [1..n] - \{j\}} a_i \cdot x_i)/a_j\Big{)}.
\]
\end{itemize}

Note that by virtue of Note \ref{note:1}
\[
D'_j = D_j \cap (b - \Sigma_{i \in [1..n] - \{j\}} int(a_i \cdot D_i))/a_j.
\]

To see that this rule preserves equivalence suppose that for some $d_1 \in D_1, \LL, d_n \in D_n$ we have
$\Sigma_{i=1}^{n} a_i \cdot d_i = b$. Then 
for $j \in [1..n]$ we have
\[
d_j = (b - \Sigma_{i \in [1..n] - \{j\}} a_i \cdot d_i)/a_j
\]
which by the Correctness Lemma \ref{lem:correctness} implies that
\[
d_j \in int\Big{(}(b - \Sigma_{i \in [1..n] - \{j\}} a_i \cdot x_i)/a_j\Big{)},
\]
i.e., $d_j \in D'_j$.

Next, consider a constraint $\Sigma_{i=1}^{n} a_i \cdot x_i \leq b$, where
$a_1, \LL ,a_n, x_1, \LL ,x_n$ and $b$ are as above.
To reason about it we can use the following rule parametrized by $j \in [1..n]$:

\begin{center}
\emph{LINEAR INEQUALITY}
\[
\frac
{\p{\Sigma_{i=1}^{n} a_i \cdot x_i \leq b}{x_1 \in D_1, \LL, x_n \in D_n}}
{\p{\Sigma_{i=1}^{n} a_i \cdot x_i \leq b}{x_1 \in D'_1, \LL, x_n \in D'_n}}
\]
\end{center}
where 

\begin{itemize}

\item for $i \neq j$
\[
D'_i := D_i,
\]

\item
\[
D'_j := D_j \cap (^{\leq} int(b - \Sigma_{i \in [1..n] - \{j\}} a_i \cdot x_i)/a_j)
\]
\end{itemize}

To see that this rule preserves equivalence suppose that for some $d_1 \in D_1, \LL, d_n \in D_n$ we have
$\Sigma_{i=1}^{n} a_i \cdot d_i \leq b$. Then
$
a_j \cdot d_j \leq b - \Sigma_{i \in [1..n] - \{j\}} a_i \cdot d_i
$.
By the Correctness Lemma \ref{lem:correctness}
\[
b - \Sigma_{i \in [1..n] - \{j\}} a_i \cdot d_i \in int(b - \Sigma_{i \in [1..n] - \{j\}} a_i \cdot x_i),
\]
so by definition
\[
a_j \cdot d_j \in ^{\leq}int(b - \Sigma_{i \in [1..n] - \{j\}} a_i \cdot x_i)
\]
and consequently
\[
d_j \in ^{\leq} int(b - \Sigma_{i \in [1..n] - \{j\}} a_i \cdot x_i)/a_j
\]
This implies that $d_j \in D'_j$.

\section{Constraint Propagation: First Approach}
\label{sec:first}
We now move on to a discussion of constraint propagation for the arithmetic constraints on integer intervals.
To illustrate the first approach consider the following example. Consider the constraint 
\[
x^3 y - x \leq 40
\]
and the ranges $x \in [1..100]$ and $y \in [1..100]$.  We can rewrite
it as 
\begin{equation}
  \label{eq:x3}
x \leq \adjfloor{\sqrt[3]{\frac{40 + x}{y}}}
\end{equation}
since $x$ assumes integer values.  The maximum value the expression on
the right-hand side can take is $\adjfloor{\sqrt[3]{140}}$, so we
  conclude $x \leq 5$.
By reusing (\ref{eq:x3}), now with the information that $x \in [1..5]$,
we conclude that the maximum value the expression on
the right-hand side of (\ref{eq:x3})
can take is actually $\adjfloor{\sqrt[3]{45}}$,
from which it follows that $x \leq 3$.

In the case of $y$ we can isolate it by rewriting the original
constraint as $y \leq \frac{40}{x^3} + \frac{1}{x^2}$ from which it
follows that $y \leq 41$, since by assumption $x \geq 1$.  So we could
reduce the domain of $x$ to $[1..3]$ and the domain of $y$ to
$[1..41]$.  This interval reduction is optimal, since $x = 1, y = 41$ and $x =
3, y = 1$ are both solutions to the original constraint $x^3 y - x \leq 40$.

More formally, we consider a polynomial constraint
$
\Sigma^{m}_{i=1} m_i = b
$
where $m > 0$, no monomial $m_i$ is an integer, the
power products of the monomials are pairwise
different and $b$ is an integer.
Suppose that $x_1, \LL, x_n$ are its variables ordered w.r.t.~
$\prec$.

Select a non-integer monomial $m_{\ell}$ and assume it is of the form
\[
a \cdot y_{1}^{n_1} \cdot \LL \cdot y_{k}^{n_k},
\]
where $k > 0$, $y_1, \LL, y_k$ are different variables ordered w.r.t.~
$\prec$,  $a$ is a non-zero integer and $n_1, \LL, n_k$ are
positive integers. So each $y_i$ variable equals to some variable in $\C{x_1, \LL, x_n}$.
Suppose that $y_p$ equals to $x_j$. We introduce the following proof rule:

\begin{center}
\emph{POLYNOMIAL EQUALITY}
\[
\frac
{\p{\Sigma^{n}_{i=1} m_i = b}{x_1 \in D_1, \LL, x_n \in D_n}}
{\p{\Sigma^{n}_{i=1} m_i = b}{x_1 \in D'_1, \LL, x_n \in D'_n}}
\]
\end{center}
where
\begin{itemize}
\item for $i \neq j$
\[
D'_i := D_i,
\]

\item 
\[
D'_{j} := int\left( D_{j} \cap \sqrt[\uproot{4}n_{p}]{int\left(({b- \Sigma_{i \in [1..m] - \{{\ell}\}} m_i})/s\right)}\ \right)
\]
and
\[
s := a \cdot y_{1}^{n_1} \cdot \LL \cdot y_{p-1}^{n_{p-1}} \cdot y_{p+1}^{n_{p+1}} \LL  \cdot y_{k}^{n_k}.
\]
\end{itemize}

To see that this rule preserves equivalence choose some $d_1 \in D_1, \LL, d_n \in D_n$.
To simplify the notation, given an extended arithmetic expression 
$t$ denote by  $t'$ the result of evaluating $t$ after 
each occurrence of a variable $x_i$ is replaced by $d_i$.

Suppose that 
$\Sigma_{i=1}^{m} m'_i = b$. Then
\[
d_j^{n_p} \cdot s' = b- \Sigma_{i \in [1..m] - \{{\ell}\}} m'_i,
\]
so by the Correctness Lemma \ref{lem:correctness} 
applied to $b- \Sigma_{i \in [1..m] - \{{\ell}\}} m'_i$ and to $s$
\[
d_j^{n_p} \in int(b- \Sigma_{i \in [1..m] - \{{\ell}\}} m_i)/int(s).
\]
Hence
\[
d_j \in \sqrt[\uproot{4}n_{p}]{int(b- \Sigma_{i \in [1..m] - \{{\ell}\}} m_i)/int(s)}
\]
and consequently
\[
d_j \in int\left( D_{j} \cap \sqrt[\uproot{4}n_{p}]{int\left(({b- \Sigma_{i \in [1..m] - \{{\ell}\}} m_i})/s\right)}\ \right)
\]
i.e., $d_j \in D'_j$.
\VV

Note that we do not apply $int(.)$ to the outcome of the root extraction
operation. For even $n_p$ this means that the second operand of the
intersection can be a union of two intervals, instead of a
single interval. To see why this is desirable, consider the constraint
$x^2-y=0$ in the presence of ranges $x \in [0..10]$, $y \in [25..100]$.
Using the $int(.)$ closure of the root extraction we would not be able
to update the lower bound of $x$ to 5.

Next, consider a polynomial constraint $ \Sigma^{m}_{i=1} m_i \leq b.
$ Below we adopt the assumptions and notation used when defining the
\emph{POLYNOMIAL EQUALITY} rule.
To formulate the appropriate rule we stipulate that for extended
arithmetic expressions $s$ and $t$
\[
int((^{\leq}s) / t)\ := \ ^{\geq}Q \cap\ ^{\leq}Q,
\]
with $Q = (^{\leq}int(s)) / int(t)$.

To reason about this constraint we use the following rule:

\begin{center}
\emph{POLYNOMIAL INEQUALITY}
\[
\frac
{\p{\Sigma^{n}_{i=1} m_i \leq b}{x_1 \in D_1, \LL, x_n \in D_n}}
{\p{\Sigma^{n}_{i=1} m_i \leq b}{x_1 \in D'_1, \LL, x_n \in D'_n}}
\]
\end{center}
where
\begin{itemize}
\item for $i \neq j$
\[
D'_i := D_i,
\]

\item 
\[
D'_{j} := int \left(D_{j} \cap \sqrt[\uproot{4}n_{p}]{int\left(^{\leq}(b- \Sigma_{i \in [1..m] - \{{\ell}\}} m_i)/s\right)}\ \right)
\]
\end{itemize}

To prove that this rule preserves equivalence choose some $d_1 \in D_1, \LL, d_n \in D_n$.
As above given an extended arithmetic expression 
$t$ we denote by  $t'$ the result of evaluating $t$ when 
each occurrence of a variable $x_i$ in $t$ is replaced by $d_i$.

Suppose that $\Sigma_{i=1}^{m} m'_i \leq b$. Then
\[
d_j^{n_p} \cdot s' \leq b- \Sigma_{i \in [1..m] - \{{\ell}\}} m'_i.
\]
By the Correctness Lemma \ref{lem:correctness} 
\[
b- \Sigma_{i \in [1..m] - \{{\ell}\}} m'_i \in int(b- \Sigma_{i \in [1..m] - \{{\ell}\}} m_i),
\] 
so by definition
\[
d_j^{n_p} \cdot s'  \in ^{\leq}int(b- \Sigma_{i \in [1..m] - \{{\ell}\}} m_i).
\]
Hence by the definition of the division operation on the sets of integers
\[
d_j^{n_p} \in ^{\leq}int(b- \Sigma_{i \in [1..m] - \{{\ell}\}} m_i)/int(s)
\]
Consequently
\[
d_j \in \sqrt[\uproot{4}n_{p}]{^{\leq}int(b- \Sigma_{i \in [1..m] - \{{\ell}\}} m_i)/int(s)}
\]
This implies that $d_j \in D'_j$.

Note that the set $^{\leq}int(b- \Sigma_{i \in [1..m] - \{{\ell}\}} m_i)$ is not an interval. 
So to properly implement this rule we need to extend the implementation of the
division operation discussed in Subsection \ref{subsec:implementation} to the case when
the numerator is an extended interval. Our implementation takes care of this.

In an optimized version of this approach we simplify the fractions of
two polynomials by splitting the division over addition and
subtraction and by dividing out common powers of variables and
greatest common divisors of the constant factors.  Subsequently,
fractions whose denominators have identical power products are added.
We used this optimization in the initial example by simplifying
$\frac{40 + x}{x^3}$ to $\frac{40}{x^3} + \frac{1}{x^2}$.  The reader
may check that without this simplification step we can only deduce that $y
\leq 43$.

To provide details of this optimization, given two monomials $s$ and $t$,
we denote by
\[
[\frac{s}{t}]
\]
the result of performing this simplification operation on $s$ and $t$.
For example, $[\frac{2 \cdot x^3 \cdot y}{4 \cdot x^2}]$ equals $\frac{x \cdot y}{2}$,
whereas $[\frac{4 \cdot x^3 \cdot y}{2 \cdot y^2}]$ equals $\frac{2 \cdot x^3}{y}$.

In this approach we assume that the domains of the variables $y_1,
\LL, y_{p-1}$, $y_{p+1}, \LL, y_n$ of $m_{\ell}$ do not contain 0. 
(One can easily show that this restriction is necessary here). 
For a monomial $s$ involving variables ranging over the integer intervals that do not
contain 0, the set $int(s)$ either contains only positive numbers or only negative numbers. 
In the first case we write $sign(s) = +$ and in the second case we write
$sign(s) = -$.

The new domain of the variable $x_j$ in the \emph{POLYNOMIAL INEQUALITY} 
rule is defined using two sequences $m_0'...m_n'$ and $s_0'...s_n'$ of 
extended arithmetic expressions such that
\[
    \frac{m_0'}{s_0'} = [ \frac{b}{s} ]\ \textrm{and}\
    \frac{m_i'}{s_i'} = - [ \frac{m_i}{s} ]\ \textrm{for $i \in [1..m]$.}
\]
Let $S := \{s_i' \mid i \in [0..m] - \{{\ell}\} \}$ and 
for an extended arithmetic expression $t \in S$ let 
$I_{t} := \{i \in [0..m] - \{{\ell}\} \mid s_i' = t\}$. We denote then by 
$p_{t}$ the polynomial $\sum_{i 
\in I_{t}}{m_i'}$. The new domains are then defined by
\[
    D'_{j} := int \left( D_{j} \cap \sqrt[\uproot{4}n_{p}]
    {^{\leq}int\left( \Sigma_{t \in S}\ \frac{p_{t}}{t}\right)}\ \right)
\]
if $sign(s) = +$, and by
\[
    D'_{j} := int \left( D_{j} \cap \sqrt[\uproot{4}n_{p}]
    {^{\geq}int\left(\Sigma_{t \in S}\ \frac{p_{t}}{t}\right)}\ \right)
\]
if $sign(s) = -$.
Here the $int(s)$ notation used in the Correctness Lemma
\ref{lem:correctness} is extended to expressions involving the division operator
on real intervals in the obvious way. We define the $int(.)$ operator applied
to a bounded set of real numbers, as produced by the division and addition
operators in the above two expressions for $D'_j$, to denote the smallest
interval of real numbers containing that set.
\section{Constraint Propagation: Second Approach}
\label{sec:second}
In this approach we limit our attention to a special type of
polynomial constraints, namely the ones of the form $s \ op \ b$,
where $s$ is a polynomial in which each variable occurs \emph{at most
  once} and where $b$ is an integer. We call such a constraint a
\emph{simple polynomial constraint}.  By introducing auxiliary
variables that are equated with appropriate monomials we can rewrite
each polynomial constraint into a sequence of simple polynomial
constraints.  This allows us also to compute the integer interval
domains of the auxiliary variable from the integer interval domains of
the original variables. We apply then to the simple polynomial
constraints the rules introduced in the previous section.

To see that the restriction to simple polynomial constraints can make
a difference consider the constraint
\[
100x \cdot y - 10y \cdot z = 212
\]
in presence of the ranges $x,y,z\in[1..9]$.
We rewrite it into the sequence
\[
u = x \cdot y, \ v = y \cdot z, \ 100u - 10v = 212
\]
where $u,v$ are auxiliary variables, each with the domain
$[1..81]$.

It is easy to check that the {\em POLYNOMIAL EQUALITY} rule introduced
in the previous section does not yield any domain reduction when
applied to the original constraint $100x \cdot y - 10y \cdot z = 212$.
In presence of the discussed optimization the domain of $x$ gets
reduced to $[1..3]$.  

However, if we repeatedly apply the {\em POLYNOMIAL EQUALITY} rule to
the simple polynomial constraint $100u - 10v = 212$, we eventually
reduce the domain of $u$ to the empty set (since this constraint has no
integer solution in the ranges $u,v \in [1..81]$) and consequently can
conclude that the original constraint $100x \cdot y - 10y \cdot z =
212$ has no solution in the ranges $x,y,z\in[1..9]$, without performing
any search.

\section{Constraint Propagation: Third Approach}
\label{sec:third}
In this approach we focus on a small set of `atomic' arithmetic constraints.
We call an arithmetic constraint \oldbfe{atomic} if it is in one of the following two forms:
\begin{itemize}
\item a linear constraint,

\item $x \cdot y = z$.

\end{itemize}

It is easy to see that using appropriate transformation rules
involving auxiliary variables we can transform each arithmetic
constraint to a sequence of atomic arithmetic constraints.  In this
transformation, as in the second approach, the auxiliary variables are
equated with monomials so we can easily compute their domains.

The transformation to atomic constraints can strengthen the reduction.
Consider for example the constraint
\[
   u \cdot x \cdot y + 1 = v \cdot x \cdot y
\]
and ranges $u \in [1..2]$, $v \in [3..4]$, and $x,y \in [1..4]$.
The first approach without optimization and the second approach
cannot find a solution without search.
If, as a first step in transforming this constraint into a linear constraint,
we introduce an auxiliary variable $w$ to replace $x \cdot y$, we are
effectively solving the constraint
\[
   u \cdot w + 1 = v \cdot w
\]
with the additional range $w \in [1..16]$, resulting in only one duplicate
occurrence of a variable instead of two.
With variable $w$ introduced
(or using the optimized version of the first approach) 
constraint propagation alone
finds the solution $u=2$, $v=3$, $x=1$, $y=1$.

We explained already in Section \ref{sec:intermezzo} how to
reason about linear constraints. (We omitted there the treatment
of the disequalities which is routine.)
Next, we focus on the reasoning for the multiplication constraint
$
x \cdot y = z
$
in presence of the non-empty ranges $x \in D_x$, $y \in D_y$ and $z
\in D_z$.  
To this end we introduce the following three domain reduction rules:

\begin{center}
\emph{MULTIPLICATION 1}
\[
\frac
{\p{x \cdot y = z}{x \in D_x, y \in D_y, z \in D_z}}
{\p{x \cdot y = z}{x \in D_x, y \in D_y, z \in D_z \cap int(D_x \cdot D_y)}}
\]
\end{center}

\II

\begin{center}
\emph{MULTIPLICATION 2}
\[
\frac
{\p{x \cdot y = z}{x \in D_x, y \in D_y, z \in D_z}}
{\p{x \cdot y = z}{x \in D_x \cap int(D_z/D_y), y \in D_y, z \in D_z}}
\]
\end{center}
\II

\begin{center}
\emph{MULTIPLICATION 3}
\[
\frac
{\p{x \cdot y = z}{x \in D_x, y \in D_y, z \in D_z}}
{\p{x \cdot y = z}{x \in D_x, y \in D_y \cap int(D_z/D_x), z \in D_z}}
\]
\end{center}

The way we defined the multiplication and the division of the integer
intervals ensures that the \emph{MULTIPLICATION} rules \emph{1,2} and
\emph{3} are equivalence preserving.  Consider for example the
\emph{MULTIPLICATION 2} rule.  Take some $a \in D_x, b \in D_y$ and $c
\in D_z$ such that $a \cdot b = c$. Then $a \in \C{x \in {\cal Z} \mid
  \te z \in D_z \te y \in D_y \: x \cdot y = z}$, so $a \in D_z/D_y$
and a fortiori $a \in int(D_z/D_y)$.  Consequently $a \in D_x \cap
int(D_z/D_y)$. This shows that the \emph{MULTIPLICATION 2} rule is
equivalence preserving.

The following example shows an interaction between all three
\emph{MULTIPLICATION} rules.

\begin{example} \label{exa:mult} 
Consider the CSP
\begin{equation}
  \label{eq:example}
{\p{x \cdot y = z}{x \in [1..20], y \in [9..11], z \in [155..161]}}.  
\end{equation}

To facilitate the reading we underline the modified domains.
An application of the \emph{MULTIPLICATION 2} rule yields
\[
{\p{x \cdot y = z}{x \in \underline{[16..16]}, y \in [9..11], z \in [155..161]}}
\]
since, as already noted in in Subsection \ref{subsec:implementation},
$[155..161]/[9..11]) = [16..16]$, and $[1..20] \cap int([16..16]) =
[16..16]$.  Applying now the \emph{MULTIPLICATION 3} rule we obtain
\[
\p{x \cdot y = z}{x \in [16..16], y \in \underline{[10..10]}, z \in [155..161]}
\]
since $[155..161]/[16..16] = [10..10]$ and $[9..11] \cap int([10..10]) = [10..10]$.
Next, by the application of the \emph{MULTIPLICATION 1} rule we obtain
\[
\p{x \cdot y = z}{x \in [16..16], y \in [10..10], z \in \underline{[160..160]}}
\]
since $[16..16] \cdot [10..10] = [160..160]$ and $[155..161] \cap int([160..160]) = [160..160]$.

So using all three multiplication rules we could solve the CSP (\ref{eq:example}). 
\HB
\end{example}

Now let us clarify why we did not define the 
division of the sets of integers $Z$ and $Y$ by
\[
Z/Y := \C{z/y \in {\cal Z} \mid y \in Y, z \in Z, y \neq 0}.
\]
The reason is that in that case for any set of integers $Z$ we would have
$Z/\C{0} = \ES$. Consequently, if we adopted this definition of the
division of the integer intervals, the resulting \emph{MULTIPLICATION}
\emph{2} and \emph{3} rules would not be anymore equivalence
preserving. Indeed, consider the CSP
\[
\p{x \cdot y = z}{x \in [-2..1], y  \in [0..0],z \in [-8..10]}.
\]
Then we would have $[-8..10]/[0..0]  = \ES$ and consequently
by the \emph{MULTIPLICATION} \emph{2} rule we could conclude
\[
\p{x \cdot y = z}{x \in \ES, y  \in [0..0],z \in [-8..10]}.
\]
So we reached an inconsistent CSP while the original CSP is consistent.

In the remainder of the paper we will also consider variants of this third
approach that allow squaring and exponentiation as atomic constraints.
For this purpose we explain the reasoning for the constraint
$
x = y^n
$
in presence of the non-empty ranges $x \in D_x$ and $y \in D_y$, and for $n>1$.
To this end we introduce the following two rules
in which to maintain the property that the domains are intervals
we use the $int(.)$ operation of Section \ref{sec:interval}:

\begin{center}
\emph{EXPONENTIATION}
\[
\frac
{\p{x = y^n}{x \in D_x, y \in D_y}}
{\p{x = y^n}{x \in D_x \cap int(D_y^n), y \in D_y}}
\]
\end{center}
\II

\begin{center}
\emph{ROOT EXTRACTION}
\[
\frac
{\p{x = y^n}{x \in D_x, y \in D_y}}
{\p{x = y^n}{x \in D_x, y \in int( D_y \cap \sqrt[n]{D_x})}}
\]
\end{center}

To prove that these rules are equivalence preserving suppose that for
some \mbox{$a\in D_x$} and $b \in D_y$ we have $a = b^n$. Then $a \in
D_y^n$, so $a \in int(D_y^n)$ and consequently $a \in D_x \cap
int(D_y^n)$.
Also $b \in \sqrt[n]{D_x}$, so
$b \in D_y \cap \sqrt[n]{D_x}$, and consequently
$b \in int( D_y \cap \sqrt[n]{D_x})$.
\II

\section{Implementation Details}
%====================
\label{sec-experiments}
In this section we describe the benchmark experiments that were performed
to compare the proposed approaches. These experiments were performed using
a single solver of the DICE (DIstributed Constraint Environment) framework.
DICE~\cite{zoeteweij03dice} is a framework for solver
cooperation, implemented using techniques from coordination programming.
It is developed around an experimental constraint solver, called OpenSolver,
which is particularly suited for coordination. 
The coordination and cooperation aspects are irrelevant from the point of view
of this paper. Relevant aspects of the OpenSolver are:
\begin{itemize}
\item It implements a branch-and-infer tree search algorithm for
  constraint solving. The inference stage corresponds to constraint
  propagation and is performed by repeated application of domain
  reduction functions (DRFs) that correspond to the domain reduction
  rules associated with the considered constraints.

   \item This algorithm is abstract in the sense that the actual functionality
      is determined by software plug-ins in a number of predefined categories.
      These categories correspond to various aspects of the abstract
      branch-and-infer tree search algorithm. Relevant categories are:
      variable domain types, domain reduction functions, schedulers that
      control the application of the DRFs, branching strategies that split
      the search tree after constraint propagation has terminated, and
      several categories corresponding to different aspects
      of a search strategy that determine how to traverse a search tree.
\end{itemize}

All experiments were performed using the \texttt{IntegerInterval} variable
domain type plug-in. Domains of this type consist of an indication
of the type of the interval (bounded, unbounded, left/right-bounded, or empty), 
and a pair of arbitrary precision integer bounds. This plug-in, and
the interval arithmetic operations on it are built using the GNU MP
library~\cite{granlund02gmp}.

The branching strategy that we used selects variables using the chronological
ordering in which the auxiliary variables come last.
The domain of the selected variable is split into two
subdomains using bisection, so the resulting search trees are binary
trees.  In all experiments we searched for all solutions, traversing
the entire search tree by means of depth-first leftmost-first
chronological backtracking.

For the experiments in this paper a DRF plug-in has been developed that
implements the domain reduction rules discussed in the previous sections.
The scheduler plug-in used in the benchmarks keeps cycling through
the sequence of DRFs, applying DRFs that have been scheduled for execution.
When a DRF is applied, and some variable domain is modified, all DRFs that
depend on these changes are scheduled for execution, including possibly the one
that has just been applied. The cycling stops when no more DRFs are scheduled
for execution, or when the domain of a variable becomes empty.

As an alternative to cycling, the scheduler can be supplied with a
\emph{schedule\/}: a sequence of indices into the sequence of DRFs.
The scheduler will then cycle through this schedule instead,
and consider DRFs for application in the specified order.
This is used in combination with the second and third approach, where
we distinguish \emph{user} constraints from the constraints that are
introduced to define the values of auxiliary variables.
Before considering for execution a DRF $f$ that is part of the implementation
of a user constraint, we make sure that all auxiliary variables that $f$
relies on are updated. For this purpose, the indices of the
DRFs that update these variables precede the index of $f$ in the schedule.
If $f$ can change the value of an auxiliary variable, its index is followed
by the indices of the DRFs that propagate back these changes to the
variables that define the value of this auxiliary variable.

For the third approach, there can be hierarchical dependencies between auxiliary
variables. Much like the HC4 algorithm of~\cite{granvilliers99revising},
the schedule specifies a bottom-up traversal of this hierarchy in a
forward evaluation phase and a top-down traversal in a backward propagation
phase before and after applying a DRF of a user constraint, respectively.
In the forward evaluation phase, the DRFs that are executed correspond to
rules \emph{MULTIPLICATION 1} and \emph{EXPONENTIATION}. The DRFs of the
backward propagation phase correspond to \emph{MULTIPLICATION 2} and \emph{3\/},
and \emph{ROOT EXTRACTION}.
It is easy to construct examples showing that the use of hierarchical
schedules can be beneficial compared to cycling through the rules.

The proposed approaches were implemented by first rewriting arithmetic constraints
to polynomial constraints, and then to
a sequence of DRFs that correspond with the rules of the approach used.
We considered the following methods:

\begin{description}
\item[1a] the first approach, discussed in Section \ref{sec:first},

\item[1b] the optimization of the first approach discussed at the end of Section \ref{sec:first}
that involves dividing out common powers of variables,

\item[2a] the second approach, discussed in Section \ref{sec:second}.
  The conversion to simple polynomial constraints is implemented
  by introducing an auxiliary variable for every non-linear monomial.
  This procedure may introduce more auxiliary variables than necessary.
  
\item[2b] an optimized version of approach 2a, where we stop
  introducing auxiliary variables as soon as the constraints contain
  no more duplicate occurrences of variables.

\item[3a] the third approach, discussed in Section \ref{sec:third},
  allowing only linear constraints and multiplication as atomic constraints.
\item[3b] idem, but also allowing $x = y^2$ as an atomic constraint.
\item[3c] idem, allowing $x = y^n$ for all $n >1$ as an atomic constraint.

\end{description}

Approaches 2 and 3 involve an extra rewrite step, where the auxiliary variables
are introduced. The resulting CSP is then rewritten according to approach 1a.
During the first rewrite step the hierarchical relations between the
auxiliary variables are recorded and the schedules are generated as a part
of the second rewrite step.
For approaches 2b and 3 the question of which auxiliary variables to introduce
is an optimization problem in itself. Some choices result in more
auxiliary variables than others. We have not treated this issue as an
optimization problem but relied on 
heuristics. We are confident that these yield a realistic implementation.
In our experiments we used the following benchmarks.

\paragraph*{Cubes}
The problem is to find all natural numbers $n \leq 1000$ that are a sum of four
different cubes, for example
\begin{displaymath}
   1^3 + 2^3 + 3^3 + 4^3 = 100.
\end{displaymath}
This problem is modeled as follows:
\begin{displaymath}
   \begin{array}{l}
   \langle
   1 \leq x_1,\ x_1 \leq x_2-1,\ x_2 \leq x_3-1,\ x_3 \leq x_4-1,\ x_4 \leq n,\\
   \phantom{\langle} x_1^3 + x_2^3 + x_3^3 + x_4^3 = n ;\
   n \in [1..1000],\ x_1,x_2,x_3,x_4 \in {\cal Z}
   \rangle
   \end{array}
\end{displaymath}

\paragraph*{Opt}
We are interested in finding a solution to the constraint
$x^3 + y^2 = z^3$ in the integer interval $[1..100000]$ for which
the value of $2x\cdot y - z$ is maximal.

\paragraph*{Fractions}
This problem is taken from~\cite{SS02}: find distinct
nonzero digits such that the following equation holds:
\begin{displaymath}
   \frac{A}{BC} + \frac{D}{EF} + \frac{G}{HI} = 1
\end{displaymath}
There is a variable for each letter. The initial domains are $[1..9]$.
To avoid symmetric solutions an ordering is imposed:
\begin{displaymath}
   \frac{A}{BC} \geq \frac{D}{EF} \geq \frac{G}{HI}
\end{displaymath}
Also two redundant constraints are added:
\begin{displaymath}
   3 \frac{A}{BC} \geq 1 \qquad \textrm{and} \qquad
   3 \frac{G}{HI} \leq 1
\end{displaymath}
Because division is not present in the arithmetic expressions, the
above constraints are multiplied by the denominators of the fractions
to obtain arithmetic constraints.

Two representations for this problem were studied: 
\begin{itemize}
\item \textit{fractions1} in which 
five constraints are used: one equality and four inequalities for
the ordering and the redundant constraints,

\item  \textit{fractions2}, used in
~\cite{SS02}, in which three auxiliary variables, $BC, EF$ and $HI$,
are introduced to simplify the arithmetic constraints:
$BC = 10B + C$, $EF = 10E+F$, and $HI = 10H+I$. 

\end{itemize}
Additionally, in both
representations, 36 disequalities $A \neq B$, $A \neq C$, ..., $H \neq I$
are used.

\paragraph*{Kyoto}
The problem\footnote{
V.~Dubrovsky and A.~Shvetsov.
\emph{Quantum} cyberteaser: May/June 1995,
\url{http://www.nsta.org/quantum/kyotoarc.asp}
} is to find the number $n$ such that
the alphanumeric equation

\begin{centering}
\begin{tabular}{llllll}
  & K & Y & O & T & O \\
  & K & Y & O & T & O \\
+ & K & Y & O & T & O \\
\hline
  & T & O & K & Y & O
\end{tabular}\\
\end{centering}
\noindent
has a solution in the base-$n$ number system. Our model uses a variable for
each letter and one variable for the base number.
The variables $K$ and $T$ may not be zero.
There is one large constraint for the addition, 6 disequalities
$K\neq Y$ ... $T \neq O$ and
four constraints stating that the
individual digits $K, Y, O, T$, are smaller than the base number. To spend some
CPU time, we searched base numbers 2..100.
\section{Results}
%================
\label{sec-results}
Tables~\ref{tab-elapsed} and~\ref{tab-numbers} compare the proposed
approaches on the problems defined in the previous section.
The first two columns of table~\ref{tab-elapsed} list
the number of variables and the DRFs that were used. Column nodes
lists the size of the search tree, including failures and solutions.
The next two columns list the number of times that a DRF was executed,
and the percentage of these activations that the domain of a variable was
actually modified. For the {\em opt} problem, the DRF that implements the
optimization is not counted, and its activation is not taken into account.
The elapsed times in the last column are the minimum times
(in seconds) recorded for 5 runs on a 1200 MHz Athlon CPU.
\begin{table}[htbp]
\begin{centering}
\begin{tabular}{lr|@{\hspace{2ex}}r@{\hspace{2ex}}r@{\hspace{2ex}}
                r@{\hspace{2ex}} r@{\hspace{2ex}}
                r@{\hspace{2ex}} r@{\hspace{2ex}}|
                 @{\hspace{2ex}}c@{\hspace{2ex}}c }
 &  & nvar & nDRF & nodes & activated & \%effective & elapsed & E & I \\
\hline
\textit{cubes}
& 1a & 5  &  14 & 167 & 1880 & 13.03 & 0.013 & + & = \\
& 2a & 9  &  22 & 167 & 2370 & 22.15 & 0.014 & + & = \\
& 3a & 13 &  34 & 359 & 4442 & 26.23 & 0.024 & - & - \\
& 3b & 13 &  34 & 227 & 3759 & 29.24 & 0.021 & = & - \\
\hline
\textit{opt}
& 1a & 4  &   7 &   115,469 &   5,186,968 & 42.16 &  22.037 & + & + \\
& 2a & 8  &  15 &   115,469 &   9,799,967 & 60.00 &  23.544 & + & + \\
& 3a & 10 &  21 &         ? &           ? &     ? &       ? & - & - \\
& 3b & 10 &  21 & 5,065,137 & 156,903,869 & 46.49 & 518.898 & - & - \\
\hline
\textit{fractions1}
& 1a & 9  & 154 & 11,289 & 1,193,579 &  3.65 & 16.586 & = & = \\
& 1b & 9  & 154 &  7,879 &   734,980 &  3.45 & 17.811 & = & = \\
& 2a & 37 & 210 & 11,289 & 1,410,436 & 23.27 &  5.575 & = & = \\
& 2b & 32 & 200 & 11,289 & 1,385,933 & 21.65 &  5.957 & = & = \\
& 3  & 43 & 208 & 11,131 & 1,426,186 & 27.76 &  5.635 & = & = \\
\hline
\textit{fractions2}
& 1a & 12 & 105 & 2,449 & 270,833 &  9.72 & 0.660 & = & = \\
& 1b & 12 & 105 &   989 &  94,894 &  9.12 & 0.538 & = & = \\
& 2a & 20 & 121 & 2,449 & 350,380 & 22.19 & 0.597 & = & = \\
& 2b & 15 & 111 & 2,449 & 301,855 & 17.51 & 0.547 & = & = \\
& 3  & 22 & 123 & 1,525 & 293,038 & 27.33 & 0.509 & = & = \\
\hline
\textit{kyoto}
& 1a & 5  &  37 & 87,085 & 3,299,736 &  6.09 & 23.680 & = & = \\
& 1b & 5  &  37 & 87,085 & 3,288,461 &  5.94 & 45.406 & + & + \\
& 2a & 13 &  53 & 87,085 & 3,781,414 & 23.03 & 11.784 & = & = \\
& 2b & 12 &  51 & 87,085 & 3,622,361 & 21.45 & 12.138 & = & = \\
& 3a & 16 &  60 & 87,087 & 4,275,930 & 26.70 & 22.538 & = & = \\
& 3b & 16 &  60 & 87,085 & 4,275,821 & 26.70 & 22.530 & = & = \\
& 3c & 16 &  59 & 87,085 & 3,746,532 & 23.26 & 10.466 & = & = \\
\\
\end{tabular}
\\
\end{centering}
\caption{Statistics and comparison with other solvers}
\label{tab-elapsed}
\end{table}

Table~\ref{tab-numbers} lists measured numbers of basic interval operations.
Note that for approach 1b, there are two versions of the division and addition
operations: one for integer intervals, and one for
intervals of reals of which the bounds are rational numbers
(marked $\mathcal{Q}$).
Columns multI and multF list the numbers of multiplications of two integer
intervals, and of an integer interval and an integer factor, respectively.
These are different operations in our implementation.
\begin{table}[htbp]
\begin{centering}
\begin{tabular}{lr|@{\hspace{2ex}}r@{\hspace{2ex}}r@{\hspace{2ex}}rlr@{\hspace{2ex}}r@{\hspace{2ex}}rlr}
 &  & root & exp & div & & multI & multF & sum & & total \\
\hline
\emph{cubes}
& 1a &      1 &      4 & 0 & &      0 & 5 & 4 & & 14 \\
& 2a & $<0.5$ & $<0.5$ & 0 & &      0 & 5 & 4 & & 9 \\
& 3a &      0 &      0 & 1 & &      1 & 6 & 5 & & 13 \\
& 3b & $<0.5$ & $<0.5$ & 1 & & $<0.5$ & 5 & 5 & & 11 \\
\hline
\emph{opt}
& 1a &  2,299 &  4,599 &  1,443 & &  1,444 &  11,064 &  5,187 & & 26,037 \\
& 2a &  1,636 &  1,538 &  2,150 & &    738 &   8,138 &  4,445 & & 18,645 \\
& 3a &      ? &      ? &      ? & &      ? &       ? &      ? & &      ? \\
& 3b & 21,066 & 18,105 & 54,171 & & 18,284 & 106,651 & 57,469 & & 275,747 \\
\hline
\emph{fractions1}
& 1a & 0 & 0 &    868 & & 28,916 & 14,238 & 13,444 & & 57,466 \\
& 1b & 0 & 0 &     51 & & 11,892 &  8,010 &  6,727 & & 29,584 \\
&    &   &   &  1,550 &$\mathcal{Q}$
                        &        &        &  1,355 &$\mathcal{Q}$
                                                     &        \\
& 2a & 0 & 0 &    734 & &    933 &  4,736 &  4,669 & & 11,071 \\
& 2b & 0 & 0 &    776 & &  1,509 &  5,292 &  5,147 & & 12,725 \\
& 3  & 0 & 0 &    693 & &    339 &  4,835 &  4,769 & & 10,636 \\
\hline
\emph{fractions2}
& 1a & 0 & 0 & 142 & & 690 & 304 & 212 & & 1,347 \\
& 1b & 0 & 0 &  19 & & 127 &  59 &  26 & &   344 \\
&    &   &   &  65 &$\mathcal{Q}$
                     &     &     &  49 &$\mathcal{Q}$
                                         &       \\
& 2a & 0 & 0 & 124 & & 149 & 138 &  94 & &   505 \\
& 2b & 0 & 0 & 124 & & 206 & 210 & 118 & &   658 \\
& 3  & 0 & 0 & 114 & &  46 & 142 & 101 & &   403 \\
\hline
\emph{kyoto}
& 1a &    735 & 11,040 & 1,963 & & 13,852 & 10,852 & 13,946 & & 52,388 \\
& 1b &    735 &  8,146 &   218 & &  8,955 & 12,516 & 10,592 & & 48,749 \\
&    &        &        & 4,310 &$\mathcal{Q}$
                                 &        &        &  3,277 &$\mathcal{Q}$
                                                              &        \\
& 2a &    383 &    759 & 1,590 & &    484 &  5,324 &  7,504 & & 16,044 \\
& 2b &    383 &    759 & 1,597 & &  1,360 &  5,756 &  8,008 & & 17,863 \\
& 3a &      0 &      0 & 1,991 & &    578 &  5,324 &  7,505 & & 15,397 \\
& 3b & $<0.5$ & $<0.5$ & 1,990 & &    578 &  5,324 &  7,504 & & 15,397 \\
& 3c &      1 &      1 & 1,554 & &    484 &  5,324 &  7,504 & & 14,868 \\
\\
\end{tabular}
\\
\end{centering}
\caption{Measured numbers (thousands) of interval operations}
\label{tab-numbers}
\end{table}
 
For the \emph{cubes} and \emph{opt} problems, the constraints are already
in simple form, so approaches 1a, 1b and 2b are identical. Also all non-linear
terms involve either a multiplication or an exponentiation, so also approaches 
2a and 3c are the same. 
The results of these experiments clearly show
the disadvantage of implementing exponentiation by means of multiplication:
the search space grows because we increase the number of variable occurrences
and lose the information that it is the same number that is being multiplied.
For \emph{opt} and approach 3a, the run did not complete within reasonable
time and was aborted.

Columns E and I of table~\ref{tab-elapsed} compare the propagation achieved
by our approaches with two other systems, respectively
\eclipse{} Version 5.6\footnote{
\eclipse{} Constraint Logic Programming System. See
\url{http://www-icparc.doc.ic.ac.uk/eclipse}
} using the ic library,
and ILOG Solver 5.1\footnote{
See \url{http://www.ilog.com}
} using type ILOINT.
For this purpose we ran the test problems without search, and compared the
results of constraint propagation. A mark `=' means that the computed domains
are the same, `+' that our approach achieved stronger propagation than
the solver that we compare with, and `-' that propagation is weaker.
For \emph{cubes}, \eclipse{} computes the same domains as those computed
according to approach 3b, so here the reduction is stronger than for 3a,
but weaker than for the other approaches.
For \emph{opt} \eclipse{} and ILOG Solver compute the same domains.
These domains are narrower than those computed according to approaches
3a and 3b, but the other approaches achieve stronger reduction.
In all other cases except for \emph{kyoto} and approach 1b the results
of all three solvers are the same.

For both representations for the fractions puzzle, the symbolic manipulation
of approach 1b is able to achieve a significant reduction of the search tree,
but this is not reflected in the timings. For \emph{fractions1\/} the
elapsed time even increases.
The reason is that computing the domain updates involves adding
intervals of real numbers.
The arithmetic operations on such intervals are more expensive than
their counterparts on integer intervals, because the bounds have to be
maintained as rational numbers.
Arithmetic operations on rational numbers are more expensive because they
involve the computation of greatest common divisors.
For \emph{kyoto} the symbolic manipulation did not reduce the size of the
search tree, so the effect is even more severe.

In general, the introduction of auxiliary variables leads to
a reduction of the number of interval operations compared to approach~1a.
The reason is that auxiliary variables prevent the evaluation of subexpressions
that did not change. This effect is strongest for \emph{fractions1}, where
the main constraint contains a large number of different power products.
Without auxiliary variables all power products are evaluated for every
\emph{POLYNOMIAL EQUALITY} rule defined by this constraint, even those
power products the variable domains of which did not change. With auxiliary variables
the intervals for such unmodified terms are available immediately, which leads
to a significant reduction of the number of interval multiplications.

The effect that stronger reduction is achieved as a result of introducing
auxiliary variables, mentioned in Section~\ref{sec:third}, is seen for
both representations of the \emph{fractions} benchmark. The effect described
in Section~\ref{sec:second} is not demonstrated by these experiments.

If we don't consider the symbolic manipulation of approach 1b,
approach 3c leads to the smallest total number of interval
operations in all cases, but the scheduling mechanism discussed in
Section~\ref{sec-experiments} is essential for a consistent good performance.
If for example the schedule is omitted for \emph{opt}, the number of interval
operations almost triples, and performance of approach 2a and 3c is then much
worse than that of approach 1a.

The total numbers of interval operations in table~\ref{tab-numbers} do not
fully explain all differences in elapsed times.
One of the reasons is that different interval operations have different costs.
Especially the preprocessing of the numerator interval for integer interval
division, discussed in Subsection~\ref{subsec:implementation}, is potentially
expensive, which may explain why for \emph{opt\/}, approach 1a runs faster than
approach 2a, even though the total number of interval operations is higher.
Among the many other factors that may be of influence,
some overhead is involved in applying a DRF, so if the number of
applications differs significantly for two experiments, this probably
influences the elapsed times as well (\emph{cubes\/}, 1a, 2a, \emph{opt\/}, 1a, 2a, \emph{fractions2\/}, 2a, 2b). The elapsed times are not the only measure
that is subject to implementation details. For example, we implemented division
by a constant interval $[-1..-1]$ as multiplication by a constant, which
is more efficient in our implementation. Such decisions are
reflected in the numbers reported in table~\ref{tab-numbers}.
\section{Discussion}
In this paper we discussed a number of approaches to constraint
propagation for arithmetic constraints on integer intervals. To assess
them we implemented them using the DICE (DIstributed Constraint
Environment) framework of \cite{zoeteweij03dice}, and compared their
performance on a number of benchmark problems. We can conclude that:
\begin{itemize}
   \item
      Implementation of exponentiation by multiplication gives weak reduction.
      In our third approach $x = y^n$ should be an
      atomic constraint.
   \item The optimization of the first approach, where common powers
      of variables are divided out, can significantly reduce the
      size of the search tree, but the resulting reduction steps
      rely heavily on the division and addition of rational numbers.
      These operations can be expected to be more expensive than their
      integer counterparts, because they involve the computation of
      greatest common divisors.
   \item Introducing auxiliary variables can be beneficial in two
      ways: it may strength\-en the propagation, as discussed in
      Sections~\ref{sec:second} and~\ref{sec:third},
      and it may prevent the evaluation of
      subexpressions the variable domains of which did not change.
   \item As a result, given a proper scheduling of the rules,
      the second and third approach perform better than the first approach
      without the optimization,
      in terms of numbers of interval operations. Actual performance
      depends on many implementation aspects. However for our test problems
      the results of variants 2a, 2b and 3c do not differ significantly.
\end{itemize}

In general, our implementation is slow compared to, for example, ILOG
Solver.  A likely cause is that we use arbitrary precision integers.
We chose this representation to avoid having to deal with overflow,
but an additional benefit is that large numbers can be represented
exactly.

A different approach would be to
use floating-point arithmetic and then round intervals inwards to the
largest enclosed integer interval. This was suggested
in~\cite{BenOld97} and implemented in for example
RealPaver\footnote{
\url{http://www.sciences.univ-nantes.fr/info/perso/permanents/granvil/realpaver/main.html}
}.
A benefit of this inward rounding approach is that all algorithms that
were developed for constraints on the reals are immediately
available. A disadvantage is that for large numbers no precise representation
exists, i.e., the interval defined by two consecutive floating-point numbers 
contains more than one integer. But it is debatable whether an exact
representation is required for such large numbers.

We realize that the current set of test problems is rather limited.
In addition to puzzles, some more complex non-linear integer optimization
problems should be studied. We plan to further evaluate the proposed approaches on
non-linear integer models for the SAT problem.
Also we would like to study the relationship
with the local consistency notions that have been defined for constraints on the
reals and give a proper characterization of the local consistencies
computed by our reduction rules.
\paragraph{Note}
This work was performed during the first author's stay at the School
of Computing of the National University of Singapore.  The work of the
second author was supported by NWO, The Netherlands
Organization for Scientific Research, under project number 612.069.003.

\bibliographystyle{plain}

\end{document}